\documentclass[a4paper,12pt]{iopart}
\usepackage{times}
\usepackage[utf8]{inputenc}
\usepackage{graphicx}
\usepackage{url}
\usepackage{cite}
\usepackage[pdftex]{color}

\begin{document} 

\date{17 Jul 2014}

\title[Reply by Pieplow and Henkel]{`Fully covariant radiation force on a polarizable particle'\\
---
Reply to the Comment by Volokitin and Persson}
%
%
\author{Carsten Henkel%
\
and Gregor Pieplow}
\address{Institute of Physics and Astronomy, University of Potsdam, 
Karl-Liebknecht-Str. 24/25, 14476 Potsdam, Germany}

\begin{abstract}
We argue that the theories of 
Volokitin and Persson [Comment arXiv:1405:2525],
of Dedkov and Kyasov [\emph{J.\ Phys.: Condens.\ Matter} {\bf 20} 
(2008) 354006],
and Pieplow and Henkel [\emph{New J.\ Phys.}\ {\bf 15} (2013) 023027]
agree on
the electromagnetic force on a small, polarizable
particle that is moving parallel to a planar, macroscopic body,
as far as the contribution of evanescent waves is concerned.
The apparent differences are discussed in detail and explained
by choices of units and integral transformations. We point out
in particular the role of Lorentz contraction in the procedure 
used by Volokitin and Persson, where a macroscopic body 
is `diluted' to get the force on a small particle. 
Differences that appear in the contribution of propagating photons
are briefly mentioned.
\end{abstract}

\vspace*{-6ex}

\submitto{\NJP}

\section{Force per particle of Volokitin and Persson}

In their Comment~\cite{Volokitin2014}, Volokitin and Persson (VP)
summarize an alternative calculation of the electromagnetic force 
on a neutral particle moving parallel to a planar half-space. 
Their approach is `macroscopic' in the sense that the starting point
are two half-spaces ($1$ and $2$, say) sliding one against the other
with arbitrary velocity \cite{Volokitin2008}.
The focus of the
present discussion is the lateral force (per unit area) given by a
component of the electromagnetic stress tensor, evaluated at the
surface of body~$1$. To arrive at the force between a single,
moving particle and a surface, the moving body~$2$ is `diluted'
by taking the limit (notation of VP, cgs units)
\begin{equation}
\epsilon_2( \omega ) - 1 \to
4 \pi n_2 \alpha( \omega )
\,,\qquad
|4 \pi n_2 \alpha( \omega )| \ll 1
\label{eq:dilute-limit}
\end{equation}
where $n_2$ is the number density of the constituent atoms (`particles'
in the following)
and $\alpha( \omega )$ their electric polarizability. The resulting
force on body~1 (in the frame where it is at rest, while body~$2$ moves
in the $x$-direction with velocity $v$) can be written as an integral
over electromagnetic waves. The focus of the discussion is the 
contribution of evanescent waves that takes the form [Eq.(27) of 
Ref.\cite{Volokitin2014}]
\begin{eqnarray}
\mbox{VP:}\qquad
f_{x}^{\rm part, ev} &=& 
- 8\pi \hbar
\int\limits_0^{\infty}\!\frac{ {\rm d}\omega }{ 2\pi }
\int\limits_{q > \omega / c}\!\frac{ {\rm d}^2q }{ (2\pi)^2 }
\frac{ q_x }{ \kappa }
{\rm e}^{ - 2 \kappa z }
\mathop{\rm Im} \alpha( \omega' )
\nonumber
\\
&&
{}
\times
\big[
N_1( \omega ) - N_2( \omega' )
\big]
\sum_{\mu}
	\phi_\mu
	\mathop{\rm Im} R_{1\mu}	
\label{eq:result-force-VP}
\end{eqnarray}
where $\omega' = \gamma ( \omega - q_x v )$ is the frequency of a
photon mode in the frame co-moving with the particle. 
We follow the
notation of Ref.\cite{Volokitin2014} except for: $k_z$ is denoted
$\kappa = (q^2 - \omega^2 / c^2)^{1/2}$;
and the Bose-Einstein distribution is written $N_i( \omega ) =
\frac12 [ \coth (\hbar \omega / 2 k_B T_i ) - 1 ]$ ($i = 1,2$)
with $T_1$ the local temperature of the body at rest, and 
$T_2$ the particle's temperature (evaluated in its co-moving frame).
The polarization-dependent weight functions and reflection amplitudes
are
\begin{eqnarray}
\phi_s &=& 
(\omega'/c)^2 + 2 \gamma^2 \beta^2 q_y^2
\frac{ \kappa^2 }{ q^2 }
\,,\qquad
R_{1s} = \frac{ \kappa - \kappa_1 }{ \kappa + \kappa_1 }
\\
\phi_p &=&
(\omega'/c)^2 + 2 \gamma^2 ( q^2 - \beta^2 q_x^2 )
\frac{ \kappa^2 }{ q^2 }
\,,\qquad
R_{1p} = \frac{ \epsilon_1 \kappa - \kappa_1 
}{ \epsilon_1 \kappa + \kappa_1 }
\label{eq:}
\end{eqnarray}
where $\beta = v / c$ and the medium propagation constant
$\kappa_1 = \sqrt{ \kappa^2 - (\epsilon_1 - 1) \omega^2 / c^2 }$.

\section{Comparison to Pieplow and Henkel}

Using the fact that $\kappa^2 = q^2 - (\omega/c)^2$,
the weight functions $\phi_\mu$
become identical to ours, Eqs.(65, 66) of 
Ref.\cite{Pieplow2013}. In the sector of evanescent waves,
$\kappa$ is real and positive so that 
\begin{equation}
\mathop{\rm Im} \frac{ R_{1\mu} \, {\rm e}^{ - 2 \kappa z } }{ \kappa }
=
\frac{  \, {\rm e}^{ - 2 \kappa z } }{ \kappa }
\mathop{\rm Im} R_{1\mu}
\label{eq:recover-im}
\end{equation}
The evanescent contribution to the friction force 
by Pieplow and Henkel (PH, Eqs.(67, 69) of Ref.\cite{Pieplow2013}) 
can therefore be written in the form
\begin{eqnarray}
\mbox{PH:}\qquad
f_{x}^{\rm part, ev} &=& 
\frac{ \hbar }{ \gamma }
\int\limits_{-\infty}^{\infty}\!\frac{ {\rm d}\omega }{ 2\pi }
\int\limits_{q > |\omega| / c}\!\frac{ {\rm d}^2q }{ (2\pi)^2 }
\frac{ q_x }{ \kappa }
{\rm e}^{ - 2 \kappa z }
\mathop{\rm Im} \alpha( \omega' )
\nonumber
\\
&&
{}
\times
\big[
N_1( \omega ) - N_2( \omega' )
\big]
\sum_{\mu}
	\phi_\mu
	\mathop{\rm Im} R_{1\mu}
\label{eq:result-force-PH}
\end{eqnarray}
The frequency integral can be reduced to the range $\omega \ge 0$ using
the fact that the integrand is even under the transformation
$(\omega, q_x) \mapsto (- \omega, -q_x)$: the expressions $q_x$,
$N_1( \omega ) - N_2( \omega' )$, 
$\mathop{\rm Im} \alpha( \omega' )$, and
$\mathop{\rm Im} R_{1\mu}$ are all odd under this transformation.%
\footnote[7]{%
To see this for $N_1( \omega ) - N_2( \omega' )$,
write it as a difference of $\coth$ functions.
For $\alpha( \omega )$ and $R_{1\mu}( \omega )$,
this is a property of Fourier transforms of real-valued response functions.
Specifically in $R_{1\mu}$, we use that $\kappa$ is real and positive 
for all $\omega, q_x$ in the evanescent sector.
The medium propagation constant is extended according to
$\kappa_1( -\omega ) = \kappa_1^*( \omega )$ (real $\omega$),
ensuring a retarded solution to the reflection and transmission problem 
for waves of negative frequencies.
}
The resulting
factor $2$ in front of $\int_0^{\infty} {\rm d}\omega$ brings 
Eq.(\ref{eq:result-force-PH}) into the form derived by
VP, \emph{except} that Eq.(\ref{eq:result-force-VP}) contains
an additional prefactor $-4\pi \gamma$. We now suggestion an explanation 
for this factor.

\paragraph{%
The minus sign} 
is due to the fact that VP calculate the force on
body~1 (it is dragged along by the moving particle), while
PH consider the force on the moving particle (a friction force).
Provided the latter is evaluated in the rest frame of body~1 (as done
by PH), the two
forces are opposite by Newton's \emph{actio = reactio}.

\paragraph{%
The factor $4\pi$}
is due to the choice of units: in the cgs units
used by VP, the displacement field in the dilute limit of body~2
is given by [see Eq.(\ref{eq:dilute-limit})]
\begin{equation}
\mbox{VP}:\qquad
{\bf D} = (1 + 4\pi n_2 \alpha) {\bf E}
\label{eq:displacement-cgs}
\end{equation}
while the same quantity is, in the units used by PH (vacuum
permittivity $\varepsilon_0 = 1$),
\begin{equation}
\mbox{PH}:\qquad
{\bf D} = ( 1 + n_2 \alpha ) {\bf E}
\label{eq:displacement-PH}
\end{equation}
The factor $4\pi$ can therefore be attributed to the different unit 
for the polarizability.

\paragraph{%
The factor $\gamma$}
is of course impossible to check by taking the 
non-relativistic limit. We suggest the following solution based on
the `dilute medium' procedure used by VP. 
The starting point is the lateral stress 
$\sigma_{xz}$ on body 1 at rest, a force per unit area.
One takes a slice of thickness ${\rm d}z$ of medium~$2$ that is
centered at a distance $z$ from body~1. This slice increases
the force on that body by an amount
\begin{equation}
{\rm d} F_x = A \, {\rm d}z \frac{ {\rm d}\sigma_{xz} }{ {\rm d}z }
\label{eq:}
\end{equation}
where $A$ is the area of the body.
In the dilute limit, forces are additive to that we convert this into 
a force per particle (in medium~2) by dividing by the number of particles 
in that slice 
\begin{equation}
f_x^{\rm part} = \frac{ {\rm d}F_x }{ {\rm d}N_2 }
=
\frac{ A\, {\rm d}z \, d \sigma_{xz} / {\rm d}z }{ A\, n \, {\rm d} z }
\label{eq:force-by-dilution}
\end{equation}
This is
the first formula in Eq.(27) of Ref.\cite{Volokitin2014}.

The key point is here: $n$ is the number density of body~2
as observed in the rest frame of body~1. This is the only
way that an observer fixed to body~1 can define a force per particle.
The density $n$ differs from
the number density in the co-moving frame due to the Lorentz-Fitzgerald 
contraction. Hence, we have
\begin{equation}
n = \gamma n_2
\,,\qquad
f_x^{\rm part} = \frac{ {\rm d}\sigma_{xz} }{ \gamma n_2 \, {\rm d}z }
\label{eq:Lorentz-contraction}
\end{equation}
where the number density in the co-moving frame is precisely the
density $n_2$ that appears in Eq.(\ref{eq:dilute-limit}) above.
Indeed, the dielectric response $\epsilon_2( \omega )$ is the one
in the rest frame of body~2, as required by the way 
VP and PH formulate the relativistic description:
the field incident on body~2 is transformed into its local 
rest-frame where $\epsilon_2( \omega )$ can be applied.
The Lorentz contraction of the particle density may be the explanation 
why Eq.(\ref{eq:result-force-VP}) is larger by a factor $\gamma$
compared to Eq.(\ref{eq:Lorentz-contraction}).

\section{Comparison to Dedkov and Kyasov}

In Eq.(13) of Ref.\cite{Dedkov2008}, Dedkov and Kyasov (DK)
give the following expression for
the evanescent contribution to the friction force on a moving
particle
\begin{eqnarray}
\mbox{DK:} \qquad 
f_{x}^{\rm part, ev} &=& 
\frac{ 16 \pi \hbar }{ \gamma }
\int\limits_{0}^{\infty}\!\frac{ {\rm d}\omega }{ 2\pi }
\int\limits_{q_x, q_y \ge 0 \atop q > \omega/c}
	\!\frac{ {\rm d}^2q }{ (2\pi)^2 }
\frac{ q_x }{ \kappa }\,
{\rm e}^{ - 2 \kappa z }
\nonumber
\\
&&
\Big\{
\mathop{\rm Im} \alpha( \omega_-' )
\left[ N_1( \omega )
- N_2( \omega_-' ) \right]
\sum_\mu 
\phi_\mu( \omega_-' )
\mathop{\rm Im} R_\mu
\nonumber
\\
&&
{} -
\mbox{$( \omega_-' \mapsto \omega_+')$}
\Big\}
\label{eq:force-DK}
\end{eqnarray}
where $\omega_\pm' = \gamma ( \omega \pm q_x v )$. 
We have used
the translation Table~\ref{t:Dedkov08b-to-Pieplow13} for the 
transcription into the notation of VP (except for $\kappa$ and
$N_i( \omega )$ as mentioned after Eq.(\ref{eq:result-force-VP})).
Note that for a fair comparison, we have neglected
the contribution from the magnetic polarizability 
$\alpha_{\rm m}$ and written
$\alpha_{\rm e} = \alpha$.
\begin{table}[hbt]
\begin{center}\footnotesize
\begin{tabular}{lllllllllll}
	& \multicolumn{2}{c}{temperatures}
	& \multicolumn{4}{c}{photon modes}
	& \multicolumn{1}{c}{occupation}
	& \multicolumn{3}{c}{polarization weights}
\\[0.5ex] \hline
\rule{0mm}{3.5ex}%
 DK~\cite{Dedkov2008} & $T_1$ & $T_2$ 
	& ${\bf k}$ & $q_0$ & $\tilde q_0$
	& $\gamma \omega^\pm$ 
	& $W( \omega / T_2, \omega^\pm \gamma / T_1 )$
	& $\gamma^2 \chi_{\rm e}^{(\pm)}( \omega, {\bf k} )$
	& $\Delta_{\rm e}$ & $\Delta_{\rm m}$
\\[0.5ex]
VP~\cite{Volokitin2014} & $T_2$ & $T_1$
	& ${\bf q}$ & $k_z,\,\kappa$ & $q_z$
	& $\omega_\pm'$ 
	& $2 [N_1( \omega ) - N_2( \omega'_\pm) ]$
	& $\phi_{p}( \omega_\pm' )$
	& $R_{1p}$ & $R_{1s}$
\end{tabular}
\end{center}
\caption[]{Translated notations from Dedkov and Kyasov~\cite{Dedkov2008}
to Volokitin and Persson~\cite{Volokitin2014}.}
\label{t:Dedkov08b-to-Pieplow13}
\end{table}
Eq.(\ref{eq:force-DK}) uses an integration range 
over only one quadrant in the ${\bf q}$-plane. Since the integrand
is even in $q_y$, a prefactor $2$ can be removed and the integral 
extended over the entire $q_y$-axis (restricted to evanescent waves,
of course).
The two lines in Eq.(\ref{eq:force-DK})
involving $\omega_-' = \omega'$ and $\omega_+'$ 
only differ by the sign of $q_x$ and can therefore be combined
into one integral over the $q_x$-axis (in the evanescent sector).
After these
manipulations, we arrive at Eq.(\ref{eq:result-force-VP}),
except for the factor $- 1/\gamma$. The minus sign is explained
as above. 
If one includes the Lorentz-contracted density
in the procedure for taking the dilute limit, as outlined above,
the formulas by VP and by DK may be brought into full agreement.

\section{Propagating sector}

VP do not discuss in their Comment the contribution from propagating
photons. A quick glance at their Eq.(22), first term, suggests
that the `dilution procedure' gives a result that is qualitatively
different. 
The rules spelled out
after Eq.(26) give to leading order
a contribution to the stress (force per area) on 
body~1 given by
\begin{eqnarray}
\sigma_{xz}^{\rm pr} &=&
- \hbar
\int\limits_0^{\infty}\!\frac{ {\rm d}\omega }{ 2\pi }
\int\limits_{q \le \omega / c}\!\frac{ {\rm d}^2q }{ (2\pi)^2 }
q_x (2 - |R_{1p}|^2 - |R_{1s}|^2)
\big[
N_1( \omega ) - N_2( \omega' )
\big]
\end{eqnarray}
Note that does not allow for a dilute limit because
it is not proportional to the density $n_2$. (It only depends on 
the temperature $T_2$ of the diluted body~2.)
A detailed comparison to the result given by
our approach would go beyond the purpose of this Reply, as there are
also physical reasons to expect a difference.%
\footnote[7]{For example, an infinitely thick half-space does not show 
any transmission for radiation emitted by body~1, while a single particle 
does. The expression $1 - |R_{2\mu}'|^2$ gives the absorption of a 
half-space and appears in the analogue of Eq.(19) of Ref.\cite{Volokitin2014}, first line,
to calculate the emission from body~2. If body~2 were a thin layer, 
however, also its transmission would appear here, and even become significant
in the dilute limit.
}

Let us compare in the following the results of DK and PH in the 
propagating sector.
Eq.(13) of 
Ref.\cite{Dedkov2008} by DK provides an integral representation
whose first line actually corresponds to a free-space (fs) contribution
(taking only the electric polarizability)
\begin{eqnarray}
\mbox{DK:} \qquad 
\left. f_{x}^{\rm part, pr} \right|_{\rm fs} &=& 
- \frac{ 4 \hbar \gamma }{ c^4 }
\int\limits_{0}^{\infty}\!\frac{ {\rm d}\omega }{ 2\pi }
\omega^4
\int\limits_{-1}^{1}\!{\rm d}x\,
x (1 + \beta x)^2
\nonumber\\
&& {} \times
\mathop{\rm Im}\alpha( \omega_1 )
\big[
N_1( \omega ) - N_2( \omega_1 )
\big]
	\label{eq:prop-fs-DK}
\end{eqnarray}
where $\omega_1 = \gamma \omega ( 1 + \beta x )$. 
The force in free space, filled
with blackbody radiation at temperature $T_1$, is apparent from
Eq.(56) in PH's Ref.\cite{Pieplow2013}. Eqs.(52, 54) in that paper
translate into the present notation as follows 
\begin{eqnarray}
\mbox{PH:}\qquad
\left. f_{x}^{\rm part, pr} \right|_{\rm fs} &=&
\frac{ 2 \hbar \gamma }{ \pi c^3 }
\int\limits_{0}^{\infty}\!\frac{ {\rm d}\omega }{ 2\pi }
\int\!\frac{ {\rm d}\Omega }{ 4\pi }
\omega q_x ( \omega - \beta q_x )^2
\nonumber\\
&& {} \times
\mathop{\rm Im}\alpha( \omega' )
\left[ N_1( \omega )
- N_2( \omega' ) \right]
\label{eq:}
\end{eqnarray}
where the symmetry manipulations mentioned 
after Eq.(\ref{eq:result-force-PH}) have been used for the $\omega$-integral.
We integrate over the directions of photon wave 
vectors (solid angle ${\rm d}\Omega$), 
their length being fixed to $\omega/c$. By rotational 
symmetry around the $x$-axis, this integral can be reduced to
(substitution $q_x = (\omega/c) x$)
\begin{eqnarray}
\mbox{PH:}\qquad
\left. f_{x}^{\rm part, pr} \right|_{\rm fs} &=&
\frac{ \hbar \gamma }{ \pi c^4 }
\int\limits_{0}^{\infty}\!\frac{ {\rm d}\omega }{ 2\pi }
\omega^4 
\int\limits_{-1}^{1}\!{\rm d}x\,
x ( 1 - \beta x )^2
\nonumber\\
&& {} \times
\mathop{\rm Im}\alpha( \omega' )
\left[ N_1( \omega )
- N_2( \omega' ) \right]
\label{eq:prop-fs-PH}
\end{eqnarray}
where now $\omega' = \gamma \omega ( 1 - \beta x )$.
Flipping the
sign of $x$, we arrive at Eq.(\ref{eq:prop-fs-DK}), up to a factor
$4\pi$ that arises again from the choice of units for the polarizability
(see above).

The surface-dependent part of DK, Eq.(13) in Ref.\cite{Dedkov2008},
involves the reflection coefficients and reads (see translation 
Table~\ref{t:Dedkov08b-to-Pieplow13}):
\begin{eqnarray}
\mbox{DK:} \qquad 
\left. f_{x}^{\rm part, pr} \right|_{\rm surf} &=& 
\frac{ 16 \pi \hbar }{ \gamma }
\int\limits_{0}^{\infty}\!\frac{ {\rm d}\omega }{ 2\pi }
\int\limits_{q_x, q_y \ge 0 \atop q \le \omega/c}
	\!\frac{ {\rm d}^2q }{ (2\pi)^2 }
\frac{ q_x }{ q_z }\,
( - \sin 2 q_z z )
\nonumber
\\
&&
{} \times
\Big\{
\mathop{\rm Im} \alpha( \omega_-' )
\big[ N_1( \omega ) - N_2( \omega_-' ) \big]
\sum_{\mu}
\phi_\mu( \omega_-' ) 
\mathop{\rm Im} R_{1\mu}
\nonumber
\\
&&
\qquad 
{} -
\mbox{$( \omega_-' \mapsto \omega_+')$}
\Big\}
\nonumber\\
&& {} +
\frac{ 16 \pi \hbar }{ \gamma }
\int\limits_{0}^{\infty}\!\frac{ {\rm d}\omega }{ 2\pi }
\int\limits_{q_x, q_y \ge 0 \atop q \le \omega/c}
	\!\frac{ {\rm d}^2q }{ (2\pi)^2 }
\frac{ q_x }{ q_z }\,
\cos( 2 q_z z )
\nonumber
\\
&&
{} \times
\Big\{
\mathop{\rm Im} \alpha( \omega_-' )
\big[ N_1( \omega ) - N_2( \omega_-' ) \big]
\sum_{\mu}
\phi_\mu( \omega_-' ) 
\mathop{\rm Re} R_{1\mu}
\nonumber
\\
&&
\qquad 
{} -
\mbox{$( \omega_-' \mapsto \omega_+')$}
\Big\}
\label{eq:force-DK-prop}
\end{eqnarray}
where the prescription $\{ R_{\rm e}^{\pm}, R_{\rm m}^{\pm}
\to \tilde R_{\rm e}^{\pm}, \tilde R_{\rm m}^{\pm} \}$ has been
applied as explained after Eq.(25) of Ref.\cite{Dedkov2008}.
We have used the notation 
$q_z = (\omega/c) [1 - (c q / \omega)^2]^{1/2}$ which is real.
We extend the ${\bf q}$ integral from one quadrant to the entire circle 
$q \le \omega/c$
using the manipulations described after Eq.(\ref{eq:force-DK}),
and get
\begin{eqnarray}
\mbox{DK:} \quad 
\left. f_{x}^{\rm part, pr} \right|_{\rm surf} &=& 
\frac{ 8 \pi \hbar }{ \gamma }
\int\limits_{0}^{\infty}\!\frac{ {\rm d}\omega }{ 2\pi }
\int\limits_{q \le \omega/c}
	\!\frac{ {\rm d}^2q }{ (2\pi)^2 }
\frac{ q_x }{ q_z }\,
\mathop{\rm Im} \alpha( \omega_-' )
\Big\{
\big[ N_1( \omega ) - N_2( \omega_-' ) \big]
\nonumber\\
&& {} \times
\sum_{\mu}
\phi_\mu( \omega_-' ) 
\big(
\mathop{\rm Re} R_{1\mu}
\cos2 q_z z 
-
\mathop{\rm Im} R_{1\mu}
\sin 2 q_z z 
\big)
\nonumber\\
\label{eq:final-form-prop-DK}
\end{eqnarray}

The result of PH can be found from Eqs.(67, 69) in Ref.\cite{Pieplow2013}
and is an integral identical to Eq.(\ref{eq:result-force-PH}),
with the ${\bf q}$-range restricted to $q \le |\omega|/c$ (propagating
waves) and the replacement
\begin{equation}
\frac{  \, {\rm e}^{ - 2 \kappa z } }{ \kappa }
\mathop{\rm Im} R_{1\mu}
\mapsto
\frac{ \mathop{\rm Re}( R_{1\mu} 
\, {\rm e}^{ 2 {\rm i} q_z z } ) }{ q_z }
\label{eq:im-r-and-propagating-photons}
\end{equation}
recalling that $q_z$ is real.
The manipulations mentioned after Eq.(\ref{eq:result-force-PH}) bring
this expression to a positive-frequency integral of the form
\begin{eqnarray}
\mbox{PH:}\qquad
\left. f_{x}^{\rm part, pr} \right|_{\rm surf} &=& 
\frac{ 2 \hbar }{ \gamma }
\int\limits_{0}^{\infty}\!\frac{ {\rm d}\omega }{ 2\pi }
\int\limits_{q \le \omega / c}\!\frac{ {\rm d}^2q }{ (2\pi)^2 }
\frac{ q_x }{ q_z }
\mathop{\rm Im} \alpha( \omega' )
\big[
N_1( \omega ) - N_2( \omega' )
\big]
\nonumber
\\
&&
{}
\times
\sum_{\mu}
	\phi_\mu
	( \mathop{\rm Re} R_{1\mu}
	\cos 2 q_z z 
	- \mathop{\rm Im} R_{1\mu}
	\sin 2 q_z z 
	)
\label{eq:force-PH-prop}
\end{eqnarray}
Up to the familiar $4\pi$, this is identical to 
Eq.(\ref{eq:final-form-prop-DK}) because $\omega_-' = \omega'$.

\section{Discussion}

The word `covariant' in the title of our paper~\cite{Pieplow2013}
may have led to the impression that this is the
only way to formulate a fully relativistic theory. This is of course
wrong: it is just a convenient formulation, and other approaches,
that do not work with 4-vectors and metric tensors etc., give
equally valid results, even for relativistic velocities. The calculations
of VP \cite{Volokitin2008} and DK \cite{Dedkov2008} are examples of these. 
The advantage of the `manifestly
covariant' formulation is that transformation properties are 
relatively easy to identify. For example, the transformation properties 
of the electromagnetic field and the polarization field both arise from
tensor fields, namely $F_{\mu\nu}$ and $M^{\mu\nu}$.

We have provided some technical details to show that DK and PH get
the same electromagnetic force for the particle+surface scenario,
as mentioned in \cite{Pieplow2013}.
The agreement holds for both propagating and evanescent waves and
for arbitrary temperatures. The approach of VP apparently differs
by a factor $-4\pi \gamma$ for evanescent waves. We have argued that
this factor disappears when the same units are used and
when the procedure of diluting the 
moving body takes into account the relativistic contraction 
of densities.


For propagating 
photons, a disagreement between VP and PH arises. We have argued 
that it is not obvious how to combine thermal equilibrium in a medium 
with the dilution procedure: indeed, as long as
body~2 is infinitely thick, there can be no contribution 
`from its back side' to the electromagnetic stress between bodies~1
and~2. It is well possible that a calculation where body~2 is 
a slab of finite thickness which is then diluted, will retrieve the 
particle+surface case in full, provided the photons incident on the
`back side' of the slab are in equilibrium in the same frame and temperature
as body~1. Otherwise a drag stress must be expected on body~1, 
similar to the
force on a particle that moves relative to the frame where a
thermal radiation field is in equilibrium \cite{Einstein1917}.

\bigskip

%

\providecommand{\newblock}{}


\begin{thebibliography}{1}
\expandafter\ifx\csname url\endcsname\relax
  \def\url#1{{\tt #1}}\fi
\expandafter\ifx\csname urlprefix\endcsname\relax\def\urlprefix{URL }\fi
\providecommand{\eprint}[2][]{\url{#2}}

\bibitem{Volokitin2014}
Volokitin A~I and Persson B~N~J 2014 Comment on "fully covariant radiation
  force on a polarizable particle" arXiv:1405.2525

\bibitem{Volokitin2008}
Volokitin A~I and Persson B~N~J 2008 {\em Phys. Rev. B\/} {\bf 78} 155437
  erratum: Phys. Rev. B {\bf 81} (2010) 23901(E)

\bibitem{Pieplow2013}
Pieplow G and Henkel C 2013 {\em New J. Phys.\/} {\bf 15} 023027 (17pp)

\bibitem{Dedkov2008}
Dedkov G~V and Kyasov A~A 2008 {\em J. Phys.: Condens. Matter\/} {\bf 20}
  354006

\bibitem{Einstein1917}
Einstein A 1917 {\em Physik. Zeitschr.\/} {\bf 18} 121--28

\end{thebibliography}

\end{document}